\documentclass{article}
\pdfoutput=1
\usepackage{spconf, amsmath, graphicx, amsfonts, mathtools}
\usepackage{epstopdf}
\usepackage{color}

\usepackage{url}
\usepackage{tikz}
\usepackage{subcaption}
\usepackage{multirow}

\DeclareMathOperator*{\argmin}{arg\,min}
\newcommand{\vW}{\mathbf{W}}
\newcommand{\vX}{\mathbf{X}}
\newcommand{\vY}{\mathbf{Y}}
\newcommand{\RR}{\mathbb{R}}
\tikzset{font=\Huge}

\title{SINGING VOICE CORRECTION 
       USING CANONICAL TIME WARPING}

\name{Yin-Jyun Luo$^{1}$, Ming-Tso Chen$^{2}$, 
Tai-Shih Chi$^{2}$, Li Su$^{1}$}
\address{$^{1}$ Institute of Information Science, Academia Sinica, Taiwan \\
    $^{2}$ Department of Electrical and Computer Engineering, National Chiao Tung University, Taiwan}
\begin{document}
%
\maketitle
\begin{abstract}
Expressive singing voice correction is an appealing but challenging problem. A robust time-warping algorithm which synchronizes two singing recordings can provide a promising solution. We thereby propose to address the problem by canonical time warping (CTW) 
which aligns amateur singing recordings to professional ones. A new pitch contour is generated given the alignment information, and a pitch-corrected singing is synthesized back through the vocoder. The objective evaluation shows that CTW is robust against pitch-shifting and time-stretching effects, and the subjective test demonstrates that CTW prevails the other methods including DTW and the commercial auto-tuning software. Finally, we demonstrate the applicability of the proposed method in a practical, real-world scenario.
\end{abstract}
\section{Introduction}
\label{sec:intro}
Synchronizing a music signal with another sequence of different modality, such as audio, \cite{montecchio2011unified}
MIDI \cite{raffel2016optimizing},
sheet music \cite{ewert2009high},
lyrics \cite{lyric2audio},
or video \cite{seeandlisten2017bochen},
provides great possibilities in automatizing music annotation \cite{thickstun2016learning},
interactive performance \cite{grachten2013automatic},
education \cite{fukuda2015score},
and many other applications. 
In this paper, we focus on the use of audio synchronization algorithms to solve the {\em singing voice correction} problem. Our main focus of singing voice correction is to automatically modify the unsatisfactory singing of an amateur singer with wrong key and intonation into correct one, while preserving original timbre as much as possible. Here the `unsatisfactory' singing, namely the {\em source}, is an audio signal, while the `correct' singing, namely the {\em target}, can be a predefined music scale \cite{melodyne}, a MIDI score of the same song \cite{azarov2014guslar}, 
a phonetic sequence of the lyric \cite{nakano2009vocalistener}, or a recording of the same song performed by a professional singer. Previous studies on this topic mostly propose rules or time-warping algorithms to shift the source pitch contour to fit a musical scale or a MIDI score, then use a vocoder to resynthesize the signal with the new pitch contour and the spectral envelope of the source \cite{azarov2014guslar}. For instance, intonation is adjusted locally using rule-based algorithms such as the dynamic pitch warping (DPW) \cite{perrotin2016target}. Since the ground truth given here are deadpan notes, the outcomes of these techniques tend to lose the voice articulation such as the sliding and vibrato, which are important for human expression. 

One way to achieve expressive singing voice correction is to align the source recording to a professional recording of the same song as the target, then modify the source pitch contour according to the alignment path. In this way, an amateur singer can get the professional singer's key, intonation, and even expression all at once. Seemingly a straightforward idea, such a task is however quite challenging because of not only the wrong pitch in the source, but also the timbre among various sources and targets. Furthermore, since a clean professional monophonic singing is seldom available, in practice one would need to consider the alignment from the user's input (typically monophonic singing) to a polyphonic target containing singing as well as accompaniment. To the best of our knowledge, no study has investigated singing voice correction in such a practical scenario.

Other studies on the alignment of singing voice, such as those for automatic singing voice assessment \cite{molina2013fundamental}, melody similarity measurement \cite{kroher2014flamenco}, speech-to-singing synthesis \cite{cen2012template}, and an adaptive karaoke system \cite{wada2017SMC} also face similar challenges caused by high-modality data, in which the conventional dynamic time warping (DTW) can not handle well. They usually focus on pertaining alignment accuracy using DTW on low-level features, but ignore the expressive parts of the features. For example, the targets for alignment are dull singings without vibrato in \cite{molina2013fundamental}. In \cite{kroher2014flamenco}, the alignment is performed on automatically transcribed notes where the vibrato and ornamentation are also suppressed. Also in \cite{wada2017SMC}, it is reported that DTW is inferior in aligning clean singings to source-separated professional ones.

These issues motivate us to investigate the use of advanced time-warping algorithms, such as canonical time warping (CTW) \cite{zhou2009canonical}, to solve the singing voice correction problem. CTW combines the canonical correlation analysis (CCA) and DTW, to offer more robustness and flexibility to cross-domain features than DTW. Although CTW has been used in the lyrics-to-audio alignment task \cite{lyric2audio}, its application in aligning an amateur singing to a professional singing with or without accompaniment, is never investigated. To verify the potential of CTW in a more accurate way, we compiled a professional-singing dataset parallel to the amateur-singing MIR-1k dataset \cite{hsu2010mir1k}. This simplifies the problem a bit to allow us to extract the target pitch contour from monophonic pitch tracking algorithm while avoiding melody tracking in polyphonic music.

The remaining part of this paper is organized as follows. The method and the proposed system are presented in Sec. \ref{sec:Methodology}. The new dataset is presented in Sec. \ref{sec:dataset}. The experiments and results are detailed in \ref{sec:exp}. We conclude the paper in Sec. \ref{sec:conclusion}.

\section{Methodology}
\label{sec:Methodology}
\begin{figure}[t]
\begin{minipage}[b]{1.0\linewidth}
  \centering
  \centerline{\includegraphics[width=8.5cm]{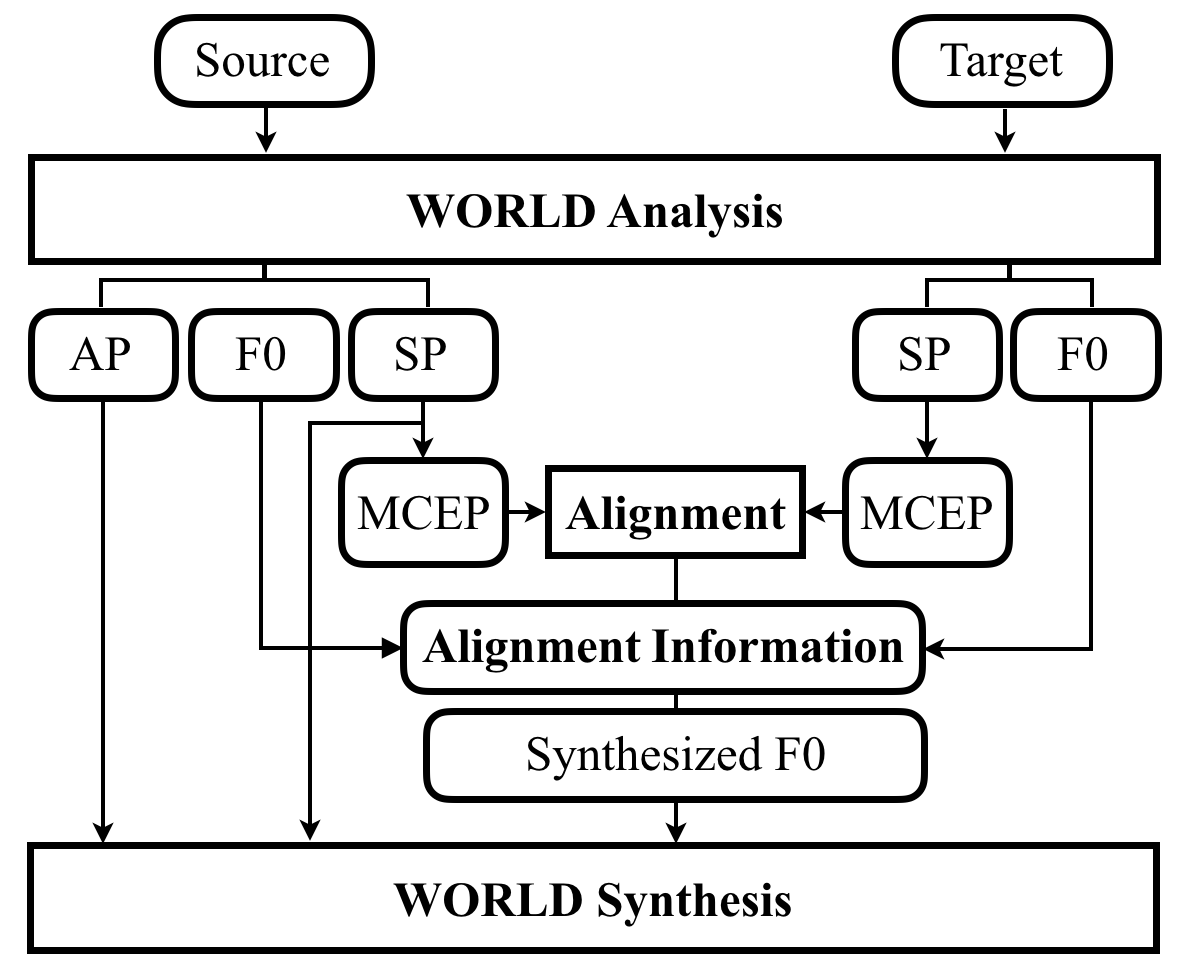}}
\end{minipage}
\caption{The proposed system flowchart}
\label{fig:flowchart}
\vspace{-4mm}
\end{figure}

Fig. \ref{fig:flowchart} illustrates the system diagram with major building blocks including feature extraction, alignment and synthesis. 

\subsection{Alignment}
\label{subsec:alignment}
Given two time sequences, $\vX\in\RR^{d \times n_x}$ and
$\vY\in\RR^{d \times n_y}$, where 
\(d\) and \(n\) denote the feature and temporal dimension, respectively. A DTW algorithm aligns the samples in \(\mathbf{X}\) and \(\mathbf{Y}\) by optimizing the following cost:
\begin{equation}
\label{eq:dtw}
\{\vW_x,\vW_y\}=\argmin_{\{\vW_x,\vW_y\}}\|\vX\vW^T_x - \vY\vW^T_y\|^2_F,
\end{equation}
where \(\mathbf{W_x}\in\{0,1\}^{m\times n_x}\) and 
\(\mathbf{W_y}\in\{0,1\}^{m\times n_y}\) are binary-valued matrices representing the aligned indexes of the two sequences, and \(m\) is the length of the warping path. Despite the efficiency of the algorithm, DTW usually falls short of aligning two sequences with nonlinear distortion between them, e.g., an in-tune singing voice and an out-of-tune one. Though being drastically different, a pair of source-target singing could share common characteristics. We thereby investigate CTW, an advanced technique that combines CCA \cite{anderson2003introduction} and DTW, to offer  feature selection and alignment at the same time \cite{zhou2009canonical}. The objective function of CTW therefore has additional linear transformation comparing to (\ref{eq:dtw}):
\begin{equation}
\label{eq:ctw}
\{\vW_x,\vW_y\}=\argmin_{\{\vW_x,\vW_y\}}\|\mathbf{V}^T_x\mathbf{X}\mathbf{W}^T_x - \mathbf{V}^T_y\mathbf{Y}\mathbf{W}^T_y\|^2_F,
\end{equation}
where \(\mathbf{V}_x\in\mathbb{R}^{d\times b}\) and \(\mathbf{V}_y\in\mathbb{R}^{d\times b}\) are the projection matrices for \(\mathbf{X}\) and \(\mathbf{Y}\), respectively, and \(b\) is the dimension of the low-dimensional embedded space sought by CCA, where \(b \leq d\). Inherited from CCA, multiple constraints are imposed for CTW, and the optimization rule can be found in \cite{zhou2009canonical}.
In order to solve Eq. (\ref{eq:ctw}), we consider initializing \(\mathbf{W}_x\) and \(\mathbf{W}_y\) in two ways. One is with DTW by solving Eq. (\ref{eq:dtw}), and the other is by linearly scaling, as in \cite{lyric2audio}, the temporal dimensions of the input sequences \(\mathbf{X}\) and \(\mathbf{Y}\) such that the two sequences share a common length \(L = \max{(n_x, n_y)}\). In this paper, we denote CTW initialized with the two methods as CTW-dtw and CTW-uniform, respectively, which will be compared in Sec. \ref{sec:exp}. 

\subsection{Singing voice analysis and synthesis}
\label{subsec:vocoder}
In this work, we leverage alignment information to achieve expressive, yet efficient, singing voice pitch correction. To analyze audio and synthesize ones from manipulated pitch contours, we use WORLD \cite{morise2016world} (D4C edition \cite{MORISE201657}) which is a vocoder-based speech analysis/synthesis system. It is shown to reduce computational cost without performance deterioration compared to STRAIGHT \cite{kawahara2008tandem} which is widely used in the literature. WORLD extracts parameters including spectral envelope (SP), fundamental frequency (F0) and aperiodicity (AP), from input audio. In this study, 1024-point FFT and 5-ms frame shift are used to extract the parameters. As the representative feature for alignment, 24-ordered mel-cepstral coefficients (MCEP) are derived for every frame from SP. Such a setting is utilized because it is common in aligning pairs of speech and for synthesis purpose in voice conversion \cite{vc2017}.
 
Note that in the subjective test, an optimal warping path is derived from either DTW or CTW given source and target audio features. Our goal is to synthesize a pitch-corrected singing with source SP, AP and modified pitch contour. We keep same the length of the modified and unmodified source pitch contour, by duplicating or averaging the F0 values in target pitch contour according to a derived warping path. SP is left untouched to preserve the timbre of source singing.

\begin{figure*}[ht!]
\includegraphics[width=\textwidth, height=5cm]{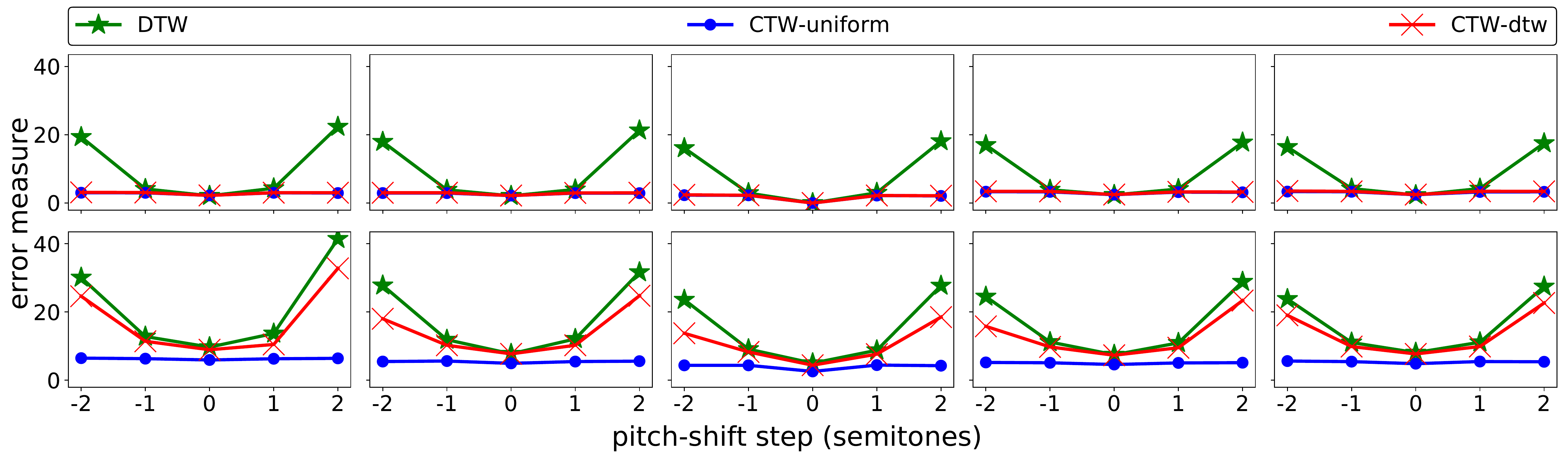}
\vspace{-6mm}
\caption{The error measure \(e\) of linear time-stretch. Top panel: \textit{mono-to-mono}; bottom panel: \textit{mono-to-poly}. From left to right: time-stretch rate \(r=0.8, 0.9, 1.0, 1.1, 1.2\).}
\label{fig:linear}
\vspace{-4mm}
\end{figure*}

\section{Dataset}
\label{sec:dataset}
For the purpose of this study, parallel songs that are sung by amateur/below-average and professional/above-average singers are necessary. Moreover, monophonic singing tracks are favored for better performance of an F0 extractor, on which synthesis quality depends. Such a parallel dataset is not within reach to our best knowledge. We therefore compile a new dataset based on the MIR-1k\footnote{the MIR-1k dataset is in \url{https://sites.google.com/site/unvoicedsoundseparation/mir-1k}.} dataset covered by amateurs as the source data, 
then recruit singers to build the target data. Although a more reasonable way is simply using the high-quality professional singings available online as our target data, we did not adopt this way because most of the state-of-the-art melody extraction algorithm from polyphonic music are still not that satisfactory for a subjective test.

Four female and five male singers were recruited to record the target singings of the dataset. Though not up to the professional level, they are with better singing skills than the ones 
in MIR-1k, and are therefore feasible to record the data to verify the performance of a singing voice correction algorithm. We picked 104 songs in MIR-1k and assigned around 10 to 30 songs to each singer depending on his or her acquaintance with the songs. The recruited singers were demanded to sing in-tune and temporally correct. The signing voice was then recorded independently as a mono-channel track. Though there were 216 recordings in total, we only selected three songs for the subjective test (see Sec. \ref{sec:subjectivetest}) to demonstrate the proposed system. On the other hand, 104 songs in the original MIR-1k are used in the objective test. For each song, only a segment of the first 30 seconds is used (see Sec. \ref{sec:objectivetest}). 

\section{Experiments and results}
\label{sec:exp}

\subsection{Objective Test}
\label{sec:objectivetest}
\begin{figure}[t]
\centering
\includegraphics[width=\linewidth, height=3cm]{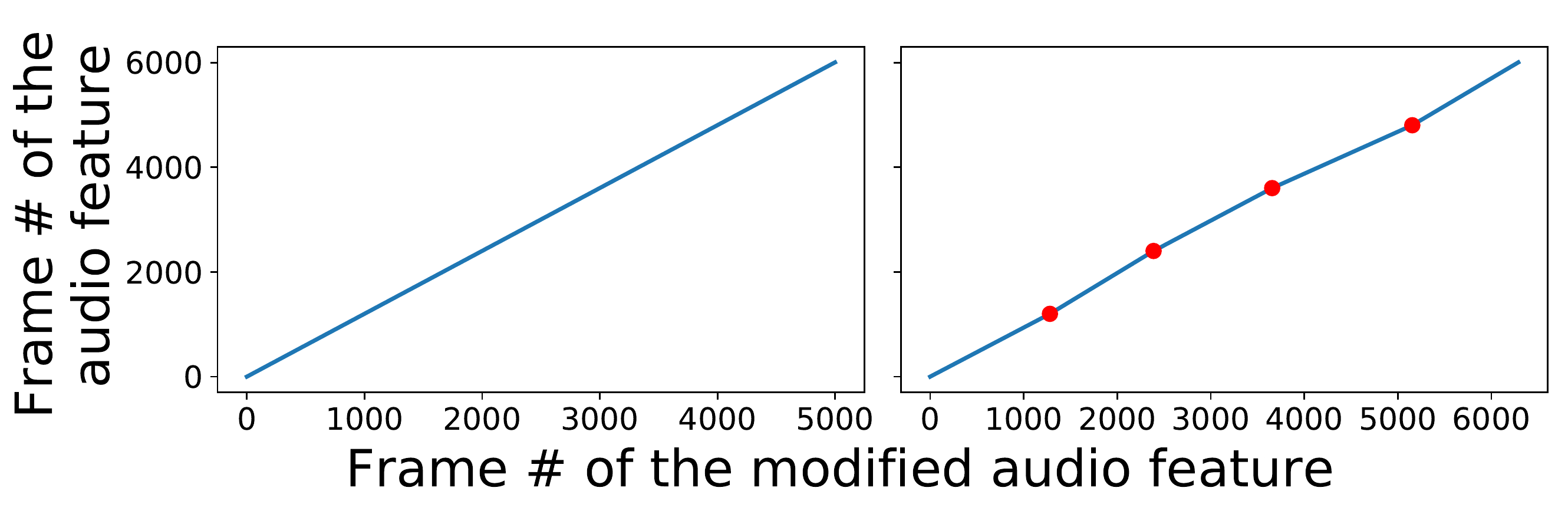}
\vspace{-6mm}
\caption{Examples of ground-truth warping paths. Left: linear, \(r=1.2\); right: nonlinear, \(r=0.93, 1.08, 0.94, 0.80, 1.06\). The dots denote the chunk boundaries.}
\vspace{-4mm}
\label{fig:nlin_wp}
\end{figure}

The three time-warping algorithms, namely DTW, CTW-uniform and CTW-dtw, are tested under two tasks. The first one, named as the {\em mono-to-mono} task, is to align two different monophonic singing voice tracks, while the second one, named {\em mono-to-poly}, is a more practical and challenging task that aims to align a singing voice track to a mixture having both singing voice and accompaniment.
Since there is no ground-truth annotation for the alignment path between a source and a target, we manage to generate sample pairs with ground truth in a synthetic way. Specifically, we employ time-stretching and pitch-shifting effects with predetermined parameters on the 104 song in the original MIR-1k, and thereby obtain the sample pairs and known alignment path. Two scenarios are considered: in the {\em linear} time-stretching scenario, five time-stretching rates \(r \in \{0.8, 0.9, 1.0, 1.1, 1.2\}\) and five pitch-shifting steps \(s \in \{-2, -1, 0, 1, 2\}\) are considered, resulting in 25 distinct parameter sets. 
In the {\em non-linear} time-stretching scenario, the same five pitch-shifting steps are also used, while every song is divided into 5 equal-length chunks, each assigned with a randomly sampled $r$ from a uniform distribution \(\mathcal{U}(0.8, 1.2)\) in order to provide irregular speed variation. Fig. \ref{fig:nlin_wp} illustrates examples of ground-truth warping paths of the linear and nonlinear scenario. The time-stretching and pitch-shifting effects are implemented by \texttt{librosa} \cite{librosa}.
We evaluate the performance by an error measure \(e\) comparing the discrepancy between the estimated warping path and the ground-truth warping path; the definition of such an error measure can be found in \cite{zhou2012generalized}. The lower the error measure is, the better the system performs.

\begin{figure}[t]
\includegraphics[width=\linewidth, height=3.5cm]{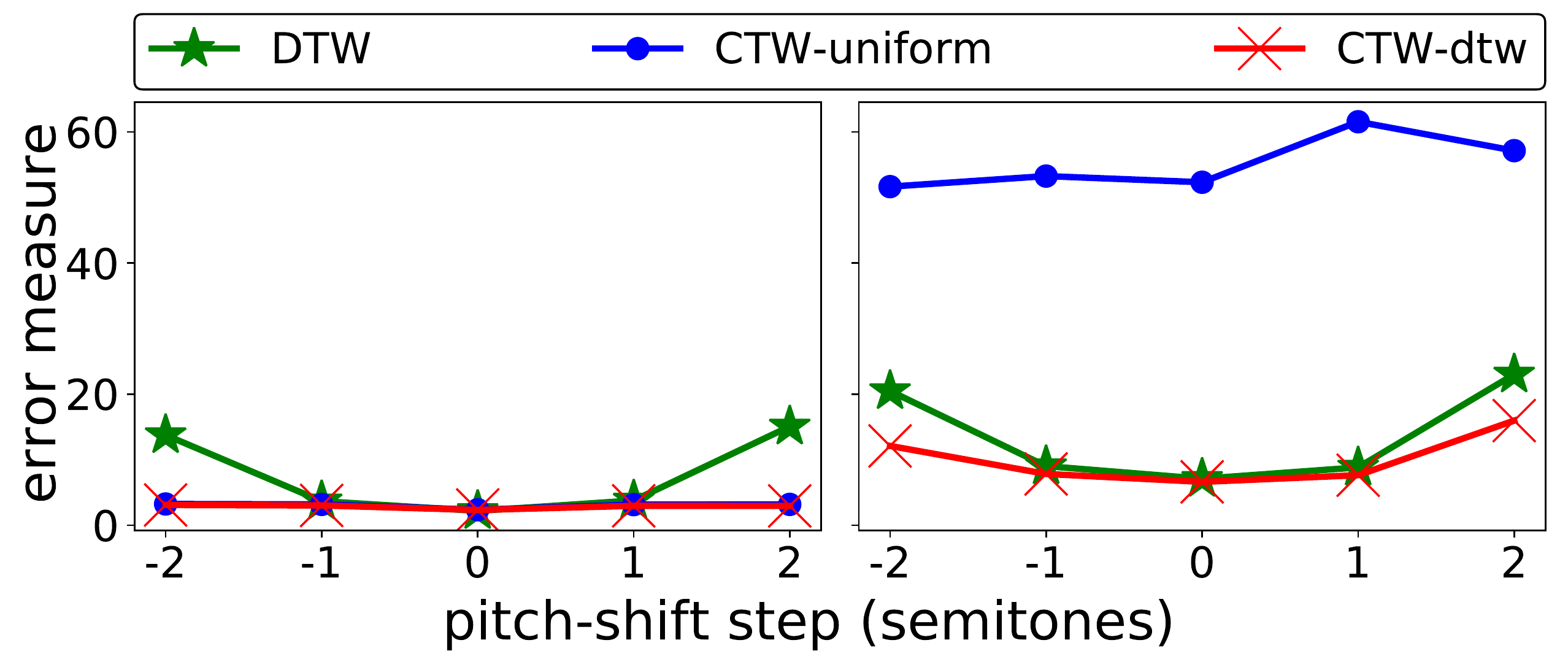}
\vspace{-6mm}
\caption{The error measure \(e\) of nonlinear time-stretch. Left: \textit{mono-to-mono}; right: \textit{mono-to-poly}.}
\label{fig:nonlinear}
\end{figure}

Fig. \ref{fig:linear} and \ref{fig:nonlinear} shows the median of the error measure $e$ over all songs in the testing dataset under each parameter set for the linear and nonlinear time-stretching scenario, respectively. 
Overall, CTW-based methods are more robust than DTW to the pitch-shifting effect, a nonlinear distortion that causes instability in the distance measure between features. It is worth noting that,  in the mono-to-poly task, CTW-uniform outperforms the other two algorithms in the linear time-stretching scenario, while CTW-dtw outperforms the others in the nonlinear ones. Moreover, CTW-uniform performs the worst under the nonlinear scenario. These results indicates the importance of initialization when performing CTW. Our results suggest that, when there is severe temporal discrepancy between two sequences, initialization with DTW can help stabilize the performance, while uniform initialization may force the CCA projects the features into a confounding subspace that simply makes the resulting features be aligned uniformly. 

\subsection{Subjective Test}
\label{sec:subjectivetest}
For subjective test, we conduct an online questionnaire\footnote{The online questionnaire can be found in \url{ https://goo.gl/forms/hXdFtUaC1Z9WLXZB2}.} containing three clips of amateur singing voice, each being modified with three pitch-corrected algorithms. The first two are the DTW and CTW-uniform with mono-to-mono setting, while the third one is the auto-tuning function of the commercial software Melodyne \cite{melodyne}. In Melodyne, we firstly make comparable the pitch level of source and target, and then each individual note is snapped to the musical scale of the original CD version. To prevent the participants from fatigue, the length of each testing clip is only 15 seconds. For each song, participants are required to listen to the amateur voice and its three corresponding pitch-corrected versions, and rate their pitch accuracy and singing fluency in a range from 1 to 5. 

Results collected from 143 valid responses, including the average rating of pitch accuracy and singing fluency, are listed in Table \ref{tab:subjective}. In average, DTW and CTW-uniform improve the pitch accuracy of the source singing voice by 1.10 and 1.37 points, respectively, while Melodyne improves it by 0.48 points. Moreover, CTW-uniform outperforms DTW significantly under the two-tailed \textit{t}-test (\(p<0.01\%, d.f.=283\)). Both alignment methods outperform Melodyne, implying that the behaviors of an unsatisfactory singing, such as time-varying tonal drift and intonation drift, are hard to be corrected without an associated music score or original singing. Meanwhile, CTW-uniform also prevails in terms of singing fluency \((p<0.1\%, d.f.=283)\), which implies that CTW-uniform is more likely to preserve human singing expression and is more natural. However, the results still show relatively low ratings overall (i.e., even the best rating is lower than 3 points), probably because the targets recordings are also not as professional as the in-market ones, and are likely to leave an unsatisfactory impression to the participants. Besides, unlike the result of the objective test, DTW in the subjective test seems to be relatively insensitive to the significant pitch level difference between the source and target singings. This might be due to that the artificial pitch-shifting algorithm tends to introduce more distortion effects than the out-of-tune but natural singing.

Last but not least, in order to demonstrate the proposed method in an even more general case, we also consider the short-time Fourier transform (STFT) being the feature for alignment, as one might exploit the alignment information for various applications other than singing voice correction, and STFT is one of the most broadly used data representations in the field. Specifically, STFT is extracted from a monophonic source and its parallel target recording mixed with accompaniment music. The FFT size and hop length are kept the same with that used for SP. The dimension of STFT is reduced from 513 to 25 by principal component analysis (PCA) before alignment. We demonstrate the results on the website.\footnote{Samples can be found in \url{https://goo.gl/MxTFyj}}

\begin{table}[t]
\centering
 \begin{tabular}[hbt!]{||c|c |cccc||} 
 \hline
 & Song & Source & DTW & CTW-u & Melo. \\ 
 \hline
 \hline
 & 1 & 1.23 & 2.16 & 2.58 & 2.06 \\ 
 \cline{2-6}
Pitch & 2 & 1.26 & 2.35 & 2.90 & 1.42\\
\cline{2-6}
accuracy  & 3 & 1.34 & 2.64 & 2.45 & 1.80\\
 \cline{2-6}
 & Avg. & 1.28 & 2.38 & 2.65 & 1.76\\
 \hline
 \hline
\multirow{4}{*}{Fluency} & 1 & 2.08 & 2.50 & 2.78 & 2.43\\ 
 \cline{2-6}
 & 2 & 2.23 & 2.37 & 2.72 & 2.01\\
 \cline{2-6}
 & 3 & 2.18 & 2.32 & 2.34 & 2.19\\
 \cline{2-6}
 & Avg. & 2.16 & 2.40 & 2.61 & 2.21\\
 \hline
\end{tabular}
\captionof{table}{Subjective test results with DTW, CTW-uniform (CTW-u) and Melodyne (Melo.) methods.}
\label{tab:subjective} 
\end{table}

\section{Conclusion and Future Work}
\label{sec:conclusion}
We demonstrate the effectiveness of utilizing CTW-based algorithms with the aid of parallel data on the singing voice correction task. With the high stability in the alignment between a singing track and a mixture, and with improved pitch accuracy and fluency in the listening test, such a method is promising for real-world application.
As future work, we will extend the study by incorporating other techniques such as melody extraction from polyphonic signals and query-by-humming algorithms, to offer facilities for retrieving melody contours from polyphonic sources which are accessible on the web. 
Furthermore, since the relation between a source-target pair is highly nonlinear, more advanced algorithms such as the deep CTW \cite{trigeorgis2016deep} might lead to further improvement.


\end{document}